\documentclass{article}
\usepackage{graphicx} % Required for inserting images
\usepackage{authblk}
\RequirePackage[margin=1in]{geometry}
\usepackage{tikz} 
\usetikzlibrary{shapes,arrows,positioning,automata,backgrounds,calc,er,patterns}
\usepackage{amssymb}
\usepackage{tikz-feynman}
\tikzfeynmanset{compat=1.0.0}
\usepackage{float}
\usepackage{amsmath}
\usepackage{hyperref}
\usepackage{physics}
\usepackage{xcolor}
\usepackage{subcaption}
\hypersetup{
    colorlinks=true,       % false: boxed links; true: colored links
    linkcolor=blue,          % color of internal links
    citecolor=blue,        % color of links to bibliography
    filecolor=blue,      % color of file links
    urlcolor=blue           % color of external links
}

\usepackage{bm}
\def\bmk{{\bm k}}

\def\bmp{{\bm p}}

\def\bmq{{\bm q}}

\title{Electron scattering and the distribution of electric charge and magnetization inside nuclei}
\author[1,2]{Alex Gnech}
\author[1,2]{Lorenzo Andreoli}
\author[3]{Graham Chambers-Wall} \author[4]{Garrett King}
\author[3,5]{Saori Pastore}
\affil[1]{Department of Physics, Old Dominion University, Norfolk, VA 23529, USA}
\affil[2]{Theory Center, Jefferson Lab, Newport News, VA 23610, USA}
\affil[3]{Department of Physics, Washington University in Saint Louis, Saint Louis, MO 63130, USA}
\affil[4]{Theoretical Division, Los Alamos National Laboratory, Los Alamos, NM 87545, USA}
\affil[5]{McDonnell Center for the Space Sciences at Washington University in St. Louis, MO 63130, USA}
\date{}
\begin{document}

\maketitle
\begin{abstract}
    How are the electric and magnetic distributions carried by protons and neutrons arranged inside an atomic nucleus?  One of the most reliable ways to answer this question is to scatter electrons from nuclei. Because the electromagnetic interaction is well understood and electron beams can be prepared and detected with high precision, electron scattering acts as a microscope that probes nuclear structure across a wide range of length scales. In this chapter we discuss how electron–nucleus scattering measurements are related to the distribution of the electric charge and magnetization inside nuclei, and what these distributions reveal about nuclear structure. We present this discussion through modern theoretical tools based on {\it ab initio} approaches, which describe nuclei as interacting many-body quantum systems, with many-nucleon interactions and electroweak currents derived from first principles.
\end{abstract}

\section{Why Use Electrons To Probe Nuclei?}

Understanding how the emergent properties of atomic nuclei arise from the underlying interactions among protons and neutrons, and ultimately from Quantum Chromodynamics (QCD), remains one of the central questions in science and serves as a key motivation for this Encyclopedia. 
In this chapter, we discuss how electromagnetic probes unveil the structure of atomic nuclei and provide insight into the underlying strong interactions among nucleons.

A natural first question is: why do we use the electromagnetic interaction to study the strong interaction? This may seem counterintuitive at first. The key point is that the electromagnetic interaction is  `well understood': the small value of the fine-structure constant $\alpha\sim 1/ 137$, which characterizes the strength of the electromagnetic interaction, allows for a perturbative treatment of electromagnetic processes, providing a systematic and reliable framework for calculating observables. Among the electromagnetic probes, electrons are a preferred choice for nuclear physicists. Their very small mass allows them to interact without significantly disturbing the nucleus, and the absence of any resolved internal structure ensures that they probe the nucleus cleanly, making electrons ideal for studying nuclear dynamics. Moreover, as charged particles, electrons can be accelerated and controlled using electromagnetic fields, and scattered electron can be detected with high accuracy, providing precise information about nuclear structure.

\begin{figure}
    \centering
    \includegraphics[width=0.5\linewidth]{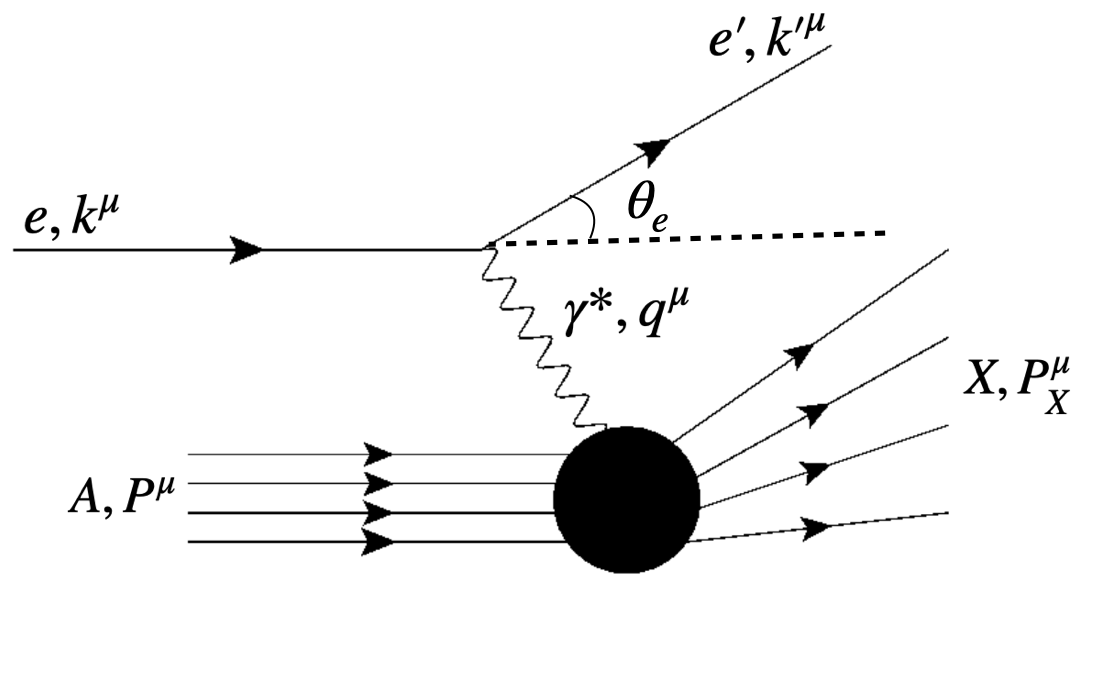}
    \caption{Schematic representation of the electron-scattering process.  The symbol $e$ indicates the incoming electron of momentum $k^\mu$, $e'$ the outgoing electron of momentum $k'^\mu$, $A$ the initial state of the nucleus with momentum $P^\mu$, and $X$ the final nuclear state with total momentum $P^\mu_X$. $\theta_e$ is the angle between the outgoing electron respect to the axis defined by the incoming electron beam.}
    \label{fig:e-scattering}
\end{figure}

In electron-scattering experiments, the primary measured quantities are the scattering cross sections, which quantify the probability that an electron is deflected by the nucleus at a given angle and energy. In this chapter we discuss the  inclusive processes, i.e. 
 \begin{equation}
     e+A\rightarrow e'+X\,,\text{  or,  } A(e,e')X\,,
 \end{equation}
 in which only the scattered electron is detected, all other final-state particles are undetected and collectively denoted by $X$.
The kinematics of the inclusive electron-scattering process is illustrated in Figure~\ref{fig:e-scattering}. In the initial state, an electron with four-momentum $k^\mu=(E_k, \bmk)$ interacts with a nucleus of mass $A$ and four-momentum $P^\mu=(E_p,\bmp)$. In first order Born approximation, the electron exchanges a single virtual photon with the nucleus, transferring energy and momentum $q^\mu=k^{\mu}-k^{\prime\mu}=(\omega,\bmq)$, where $k^{\prime\mu}=(E_{k'}, \bmk^\prime)$ is the four-momentum of the electron in the final state. In an inclusive process, only the energy, the momentum  and the angle $\theta_e$ of the emitted electron $e'$ with respect to the incoming electron beam are measured. The final nuclear states ($X$), that carry a total four-momentum $P_X^\mu=(E^X,\bmp^X)$, are  not detected. 

 The resulting cross section depends only on two independent quantities: the energy transfer $\omega$,  and the momentum transfer $q$ that can be obtained from the four-momentum and the angle $\theta_e$ of the emitted electron. 
 The energy transfer $\omega$ characterizes how the nucleus responds in time to an external probe showing the ``resonating" frequencies of the system as a whole and of the single components (the nucleons). 
The momentum transferred $|\bmq|=|\bmk-\bmk'|$  represents instead the inverse wavelength of the exchanged virtual photon, and sets the spatial resolution of the experiment. To resolve internucleon distances ranging from $\sim0.25$ fm (partially overlapping nucleons) up to  $\sim 5$ fm (where nucleons are separated  but still interact via the strong force), we require momentum transfer spanning from 
\begin{equation}
    q \equiv |\bmq|=\frac{\hbar c}{\lambda c}\sim 0.04\, \, \text{to}\,\, 0.8 \text{ GeV}/c\,,
\end{equation}
as implied by the de Broglie relation. To probe these momentum transfers, experiments use spectrometers to select electrons with specific momenta. By measuring the scattering at different momenta and angles, they reveal the distribution of the electric charge and magnetization inside the nucleus, with accelerated electrons acting as a `microscope' into the internal structure of nuclei across a wide range of length scales.

\begin{figure}
    \centering
    \includegraphics[width=0.7\linewidth]{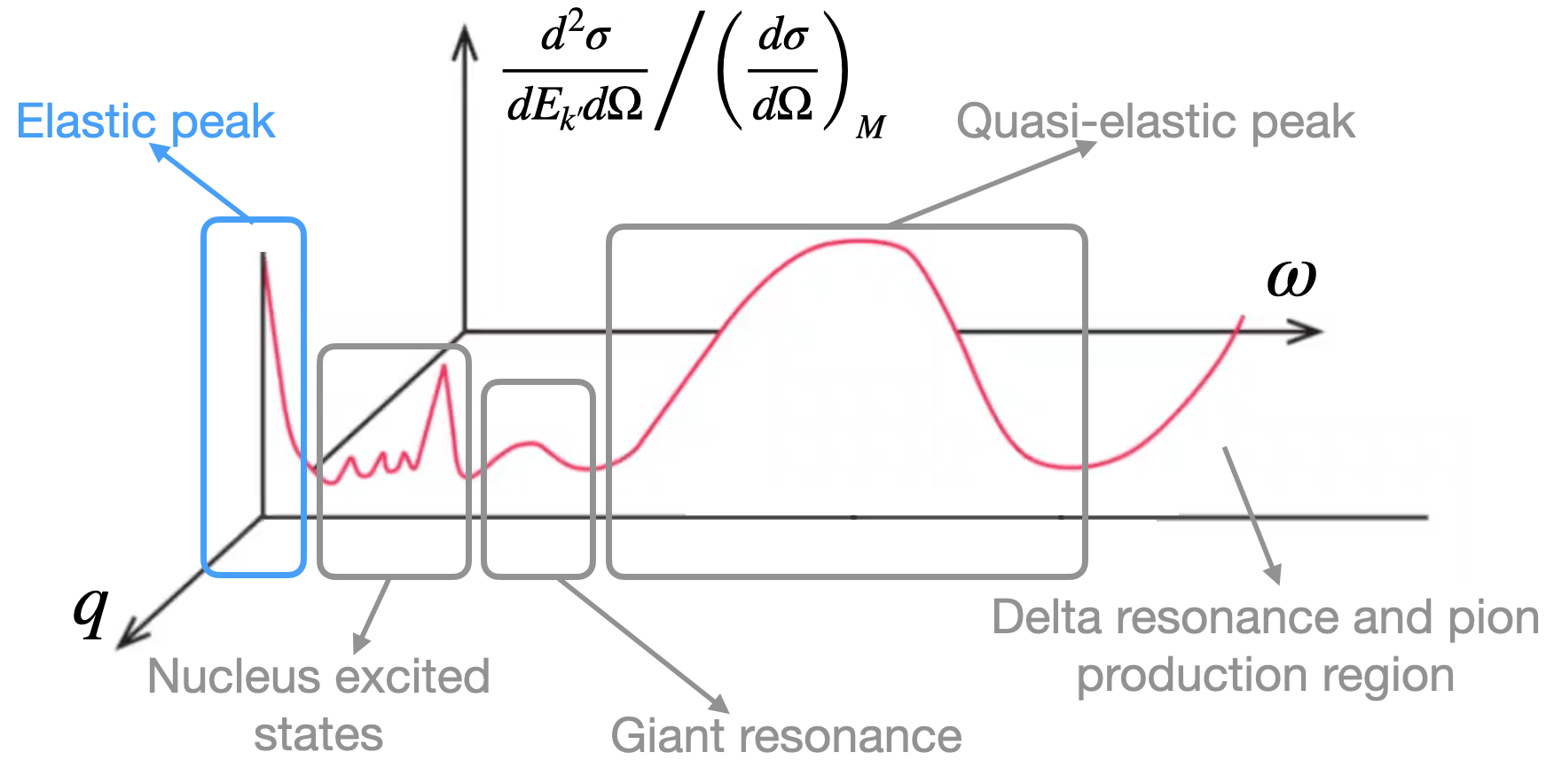}
    \caption{Pictorial representation of a typical inclusive electron-scattering cross-section (normalized to the Mott cross-section) at fixed momentum transfer $q$, as a function of the energy transfer $\omega$. The figure highlights kinematical regions corresponding to various nuclear phenomena discussed in the text. The blue frame indicate the elastic peak that correspond to the kinematic region from which the nuclear form factors are obtained (Figure adapted from Ref.~\cite{Bacca:2014}).}
    \label{fig:xsec_all}
\end{figure}

A typical inclusive cross section at fixed momentum transfer $q$ as a function of the energy transfer $\omega$, is shown schematically in Figure~\ref{fig:xsec_all}, exhibits different regions corresponding to various reaction mechanisms (``resonating" frequencies) induced by the probe. The elastic peak, at the lowest energy transfer, indicates that the nucleus remains in its ground state and provides information on the static distributions of the charge and magnetization. At higher energy transfers, the nucleus can be excited, producing discrete peaks or collective motions (giant resonances), and eventually entering the quasi-elastic region where the probe interacts mainly with a single nucleon. Beyond the quasi-elastic region, the energy transfer is sufficient to excite nucleonic resonances and induce meson production. In this chapter, we focus on the elastic scattering at $\omega\sim0$, which probes the static structure of the nucleus. 

In closing, it is worth pointing out that electron scattering provides much more information than what is achieved by the elastic scattering alone. Exclusive processes, in which additional particles in the final state are detected, provide a breadth of information about nuclear dynamics, but are theoretically challenging because all the final states must be resolved in detail. Similarly, at energy transfers beyond the quasi-elastic region, where non-nucleonic states are produced, the electron-nucleus interaction must be described using theories that incorporate degrees of freedom beyond nucleons.

This chapter is organized as follows: in Section~\ref{sec:el_xsec}, we present the cross sections for elastic electron-nucleus scattering and discuss how it provides access to the charge and magnetic distributions of the nucleus. In Section~\ref{sec:currents}, we describe the theoretical framework adopted to calculate the cross sections. Section~\ref{sec:em_ff} examines the form factors of selected light nuclei. In Section~\ref{sec:conclusions}, we summarize the discussion and draw conclusions.

\section{Electron-Nucleus Elastic  Cross Section}\label{sec:el_xsec}

In elastic scattering processes, the nucleus remains in its ground state after the interaction, \textit{i.e.},
\begin{equation}
    e+A\rightarrow e'+A\,.
\end{equation}
This means that the energy transfer to the nucleus is below the threshold for excitation or breakup, so the nucleus is not internally excited and no additional particles are produced. Elastic events can be selected experimentally through their kinematical signature. In the laboratory frame, where the target nucleus of mass $m_A$ is at rest, the energy and momentum transfer are related by
\begin{equation}\label{eq:omegael}
\omega_{el}=\sqrt{q^2+m_A^2}-m_A\,.
\end{equation}

The elastic differential cross section can be written as a function of \textit{nuclear form factors} as~\cite{Walecka:1995mi, Carlson:2014vla}:
\begin{equation}\label{eq:xsec_elastic}
\frac{d \sigma}{d \Omega_e}=4 \pi \left(\frac{d\sigma}{d\Omega_e}\right)_M f_{\mathrm{rec}}^{-1}\left[\frac{Q^4}{q^4} F_L^2(q)+\left(\tan^2\frac{\theta_e}{2}+\frac{Q^2}{2 q^2}\right) F_T^2(q)\right]\,,
\end{equation}
where $Q^2\equiv-q^\mu q_\mu = q^2-\omega^2$ is the four-momentum transfer squared,  $\Omega$ is the solid angle of the detected electron, and 
\begin{equation}
f_{rec} = 1 + \frac{2\,E_k}{m_A}  \sin^2\left(\frac{\theta_e}{2}\right)\, ,
\end{equation}
is the recoil correction, with $E_k$ incoming electron energy and $\theta_e$ the scattering angle (see Figure~\ref{fig:e-scattering}). 

\begin{figure}
    \centering
    \includegraphics[width=0.7\linewidth]{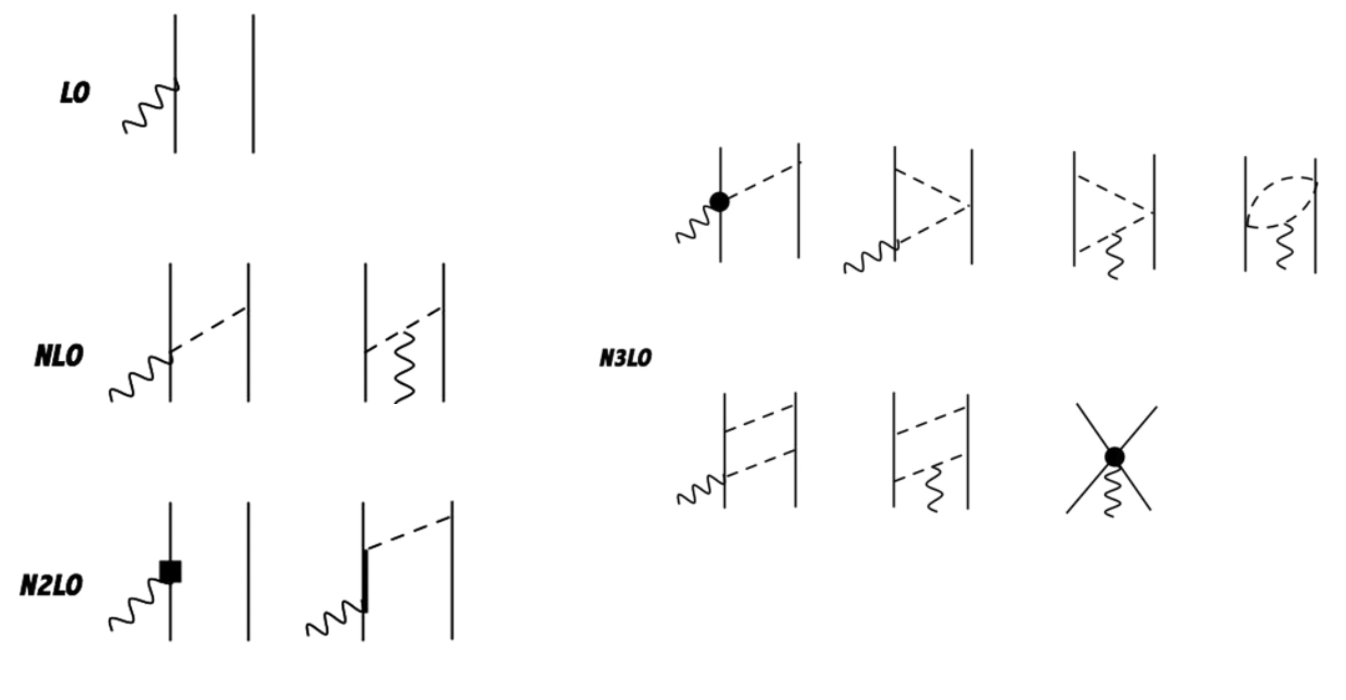}
    \caption{$\chi$EFT diagrams contributing to the electromagnetic currents $\bm j(\bm q)$ up to next-to-next-to-next-to leading order (N3LO) in chiral effective field theory. Diagrams are listed by chiral order. The full lines represent the nucleons, the dashed lines the pions, the full thick one the $\Delta$ isobar, and the wavy ones the photons.  The square represent relativistic corrections to the leading order (LO) vertex, while the dots at N3LO are associated with unknown low-energy constants. The cross diagram at N3LO is known as a contact term.  (Figure adapted from Ref.~\cite{Schiavilla:2018udt}).}
    \label{fig:em_diagrams}
\end{figure}

The factor $\left(\frac{d\sigma}{d\Omega_e}\right)_M$ is the Mott cross-section describing the scattering from a structureless, point-like particle via the Coulomb interaction:
\begin{equation}
    \left(\frac{d\sigma}{d\Omega_e}\right)_M = \left[\frac{\alpha \cos \left(\theta_e / 2\right)}{2 E_{k^\prime} \sin ^{2}\left(\theta_e / 2\right)}\right]^{2} \, ,
\end{equation}
where $E_{k'}$ is the energy of the outgoing electron and $\alpha$ is the fine-structure constant. Deviations from point-like scattering are encoded in the \textit{nuclear form factors}, which contain information about the spatial distributions of charge and magnetization inside the nucleus. 

Specifically, the \textit{longitudinal}, $F_L(q)$,  and \textit{transverse}, $F_T(q)$ form factors are defined as: 
\begin{equation}\label{eq:ff_l}
    F^2_L(q)=\frac{1}{4\pi}\frac{1}{2J_A+1}\sum_{M_A,M_A'}|\bra{\Psi(M_A)}\rho(\bmq)\ket{\Psi(M_A')}|^2\,,
\end{equation}
and 
\begin{equation}\label{eq:ff_t}
    F^2_T(q)=\frac{1}{4\pi}\frac{1}{2J_A+1}\sum_{M_A,M_A'}|\bra{\Psi(M_A)}j_x(\bmq)\ket{\Psi(M_A')}|^2+|\bra{\Psi(M_A)}j_y(\bmq)\ket{\Psi(M_A')}|^2\,,
\end{equation}
where $\Psi(M_A)$ is the wave function describing the ground state of nucleus with mass number $A$ and total angular momentum $J_A$, and projection along the $z$-axis $M_A$.  The operators $\rho(q)$ and $\bf{j}(\bmq)$ are nuclear electromagnetic charge and current operators describing how the electron couples to the nuclear electromagnetic charges. It is clear then that the longitudinal response provides information  on how the electric charges are statically distributed in the nucleus depending on the charge operator only. The situation on the transverse response is more involved and we need more information to understand its meaning. However, we anticipate here that it provides information on the spatial distribution of the magnetic charges.

The structure of the elastic cross section in terms of the two form factors can be easily derived using Lorentz invariance. The cross section appearing in Eq.~\eqref{eq:xsec_elastic} can be rewritten in a Lorentz covariant form as
\begin{equation}
    \frac{d \sigma}{d \Omega_e}\propto L_{\mu\nu}H^{\mu\nu}\,,
\end{equation}
where $H_{\mu\nu}$ is the hadronic tensor that describes the lower vertex of Figure~\ref{fig:e-scattering} and  $L_{\mu\nu}$ is the leptonic tensor that describes the upper one.
Since the only independent external four-vectors are $P^\mu$ and $k^{\prime\mu}$ ($P^\mu_X$ is fixed by conservation of energy and momentum, and $k^{\mu}$ by the initial conditions), the hadronic tensor admits a unique expansion in terms of  combinations of these two quantities. Using  conservation of current and discrete symmetries, only two independent combinations can be written, each associated with two independent scalar functions that correspond exactly to the nuclear form factors~\cite{Walecka:1995mi}. 
Being scalar functions, they can depend only on the scalar variables $\omega$ and $q$. In the elastic case, in which  $\omega$ is  constrained as in  Eq.~\eqref{eq:omegael}, the form factors  explicitly depend only on  the  momentum transfer $q$, as shown in Eq.~\eqref{eq:xsec_elastic}. 

From the experimental point of view, the longitudinal and transverse form factors can be extracted from the measured cross sections using the Rosenbluth separation, a method in which cross sections are measured at different electron scattering angles and at fixed  momentum transfer~\cite{Donnelly:1984rg}. This procedure allows to disentangle the two contributions, which is particularly important because the transverse form factor, as we will see later, is typically much smaller than the longitudinal one.

Now that we have both the theoretical framework to calculate the cross sections and the experimental measurements, we can compare our predictions with the data and see how well our models explain the electric and magnetic charge distribution of the nucleus.

\begin{figure}
    \centering
    \includegraphics[scale=1.3]{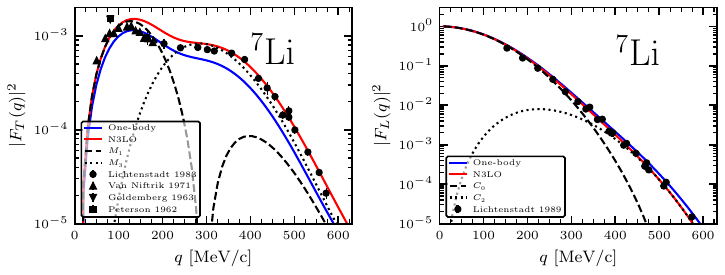}
    \caption{Magnetic (left) and charge (right) form factors computed using VMC in the $\chi$EFT framework for $^7$Li. The red (blue) lines correspond to the calculation using the full N3LO (one-body only) currents. The black dashed and dotted lines correspond to the $M_1$ and $M_3$ multipoles for the transverse form factor and $C_0$ and $C_2$ for the longitudinal form factors.
    The black dots are the experimental data taken from Refs.~\cite{Peterson1962,Goldemberg1963,Vanniftrik1971,Lichtenstadt1983} and Ref.~\cite{Lichtenstadt:1989nr} for the magnetic and charge form factors respectively.
    Figures adapted from references~\cite{Chambers-Wall:2024fha, Chambers-Wall:2024uhq,King:2024jiq}.}
    \label{fig:ffactor7li}
\end{figure}

\section{Ab Initio Approaches to the Elastic Form Factors}\label{sec:currents}

We now turn to the theoretical description of the nucleus needed to compute the form factors. Modern theories treat the nucleus as a quantum many-body system of strongly correlated nucleons, whose collective properties emerge from their interactions. Many-nucleon correlations are essential to explain basic nuclear observables, such as the energy spectra, and, similarly, many-nucleon electromagnetic currents are required to accurately describe nuclear electromagnetic properties~\cite{Hergert:2020bxy}. 

\textit{Ab initio} nuclear theories aim to retain the effects of these complex many-nucleon dynamics by solving the nuclear many-body Schr\"odinger equation with two- and, when relevant, three-nucleon forces. Modern approaches often adopt effective field theory to construct many-nucleon operators. In principle, since nucleons have internal structure and are made of interacting quarks and gluons, one should describe nuclear dynamics in terms of these fundamental degrees of freedom using quantum chromodynamics (QCD). However, this is challenging due to the non-perturbative nature of QCD at low energies. Non-perturbative methods, such as lattice QCD, have been extensively developed to address this regime, but they have been, to date, applied to study few-nucleon systems only. 

An alternative approach is to adopt the chiral effective field theory ($\chi$EFT)~\cite{Rev_EFT}, a low-energy approximation of QCD. This provides an effective description of nuclear interactions and currents in terms of nucleonic and pionic degrees of freedom while preserving the symmetries exhibited by QCD. The remaining degrees of freedom are excluded from the theory (we say ``integrated out'')  and included effectively through low-energy constants (LECs) that are typically fitted to describe the data associated to few-body systems.
This approach also allows interactions and currents to be systematically organized in a controlled  expansion in the low-momentum characteristic of the nuclear physics domain. The expansion in low-momentum gives us, in principle, a way to improve systematically the theory by computing the next order in the expansion and to obtain a theoretical error associated with the missing terms. A large portion of this Encyclopedia is dedicated to the derivation of the nuclear Hamiltonian and the computational methods used to solve the associated many-body Schr\"odinger equation; we refer the reader to those chapters for further details. 

Here, we briefly sketch the structure of many-nucleon charge and current operators to allow us to better interpret the result shown in the next section.
The charge and current operators can be written as:
\begin{align}
    \rho(\boldsymbol{q})&=\sum_i\rho_i(\boldsymbol{q})+\sum_{i<j}\rho_{ij}(\boldsymbol{q})+\ldots
    \\
    \boldsymbol{j}(\boldsymbol{q})&=\sum_i\boldsymbol{j}_i(\boldsymbol{q})+\sum_{i<j}\boldsymbol{j}_{ij}(\boldsymbol{q})+\ldots
\end{align}
where the ellipsis denotes terms with more than two nucleons. One-nucleon operators describe the interaction of the probe with individual nucleons leading to the so called Impulse Approximation. For example, the one-body charge operator describes the interaction of the external probe with the nucleonic charge distribution. In \textit{ab initio} theories, the intrinsic properties of the nucleons, including their charge distributions, are inputs taken from experiments, where available, or theoretical determinations. 

Two- and higher-body operators encode correlations among nucleons. In particular, two-nucleon currents often arise from the exchange of virtual pions between correlated nucleons. Because pions can carry electric charge, the electromagnetic probe interacts not only with individual nucleons but also with pairs of correlated nucleons through so-called meson-exchange currents, where the probe interacts with the exchanged mesons. 

In Figure~\ref{fig:em_diagrams}, we show a typical expansion of the vector electromagnetic currents derived within a chiral effective field theory in which pions, nucleons, and $\Delta$-isobars are used as explicit degrees of freedom. At  leading order (LO), the only contribution to the electromagnetic current arises from the interaction of the photon with a single nucleon. At next-to leading order (NLO), the photon interact simultaneously with two nucleons exchanging a pion. These contributions are known as pion-exchange currents. At next-to-next-to leading order (N2LO), the currents receive contributions from relativistic corrections to the impulse-approximation diagram (black square) and from the excitation of a $\Delta$-isobar excited by the photon (this diagram appears only if the $\Delta$ is included explicitly in the $\chi$EFT, otherwise it is embedded into the LECs of the $\Delta$-less theory). Finally, at next-to-next-to-next-to leading order (N3LO), there are  two-pion-exchange diagrams and short-range contact terms that encode the interaction of heavier degrees of freedom with the external photon (cross term in Figure~\ref{fig:em_diagrams}). The black dots appearing in the diagrams at this order indicate that these diagrams are associated with unknown LECs that must be determined by fitting to experimental data from reactions involving photon exchanges. The electromagnetic charge and current operators  have been derived in $\chi$EFT from different groups using different approaches. For details, we refer the reader to the following references ~\cite{Pastore2008,Pastore2009,Pastore2011,Epelbaum2009,Epelbaum2011,Krebs2019}.

With the electromagnetic currents discussed above and the wave function computed using the Hamiltonian consistent with the currents, it is then possible to compute the matrix elements shown in Eqs.~\eqref{eq:ff_l} and~\eqref{eq:ff_t} and compare the $\chi$EFT predictions with the experimental data, as discussed in the next Section. 

\begin{figure}
    \centering
    \includegraphics[scale=1.3]{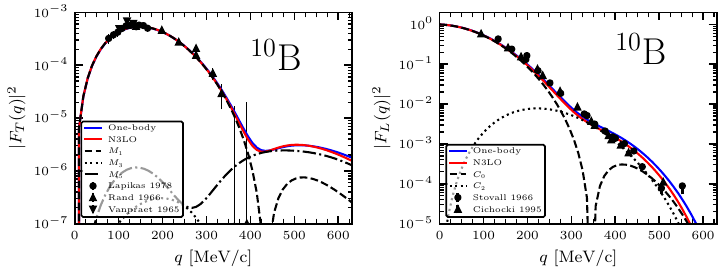}
    \caption{Magnetic (left) and charge (right) form factors computed using VMC in the $\chi$EFT framework for $^{10}$B. The red (blue) lines correspond to the calculation using the full N3LO (one-body only) currents. The black dashed, dotted and dash-dotted lines correspond to the $M_1$, $M_3$ , and $M_5$ multipoles for the transverse form factor and $C_0$ and $C_2$ for the longitudinal form factors.
    The black dots are the experimental data taken from Refs.~\cite{Lapikas1978,Rand1966,Vanpraet1965} and Refs.~\cite{Stovall:1966,Cichocki:1995zz} for the magnetic and charge form factors respectively.
    Figures adapted from references~\cite{Chambers-Wall:2024fha, Chambers-Wall:2024uhq,King:2024jiq}.}
    \label{fig:ffactor10b}
\end{figure}

\section{What form factors tell us about nuclear structure}\label{sec:em_ff}

The electromagnetic form factors carry information on how the nucleons are distributed and correlated inside a nucleus. In this section,  we make this connection  explicit by taking as illustrative examples several light nuclei recently studied using the variational Monte Carlo method (VMC)~\cite{Chambers-Wall:2024fha,Chambers-Wall:2024uhq,King:2024jiq}.

 The matrix elements in Eqs.~\eqref{eq:ff_l} and~\eqref{eq:ff_t} are multidimensional integrals that must be evaluated for each momentum value $q$, and for each combination of $M_A$ and $M_A'$. This is computationally demanding. It is therefore convenient to use a different representation of the operators based on the so called {\it multipole expansion}. The basic idea is that operators $\rho(\bmq)$ and $\boldsymbol{j}(\bmq)$ can be expanded in partial waves, from which new operators, known as multipole operators, with defined angular momentum $L$ can be constructed. The derivation is somewhat technical and can be found in Ref.~\cite{Walecka:1995mi,Forest1966}. Here, we report only the final expressions of the form factors rewritten in terms of the multipoles: 
\begin{align}\label{eq:ff_rme}
    F^2_L(q)&=\frac{1}{2J_A+1}\sum_{L\geq0}|\bra{\Psi_0}|C_L(q)|\ket{\Psi_0}|^2\,,\\
    F^2_T(q)&=\frac{1}{2J_A+1}\sum_{L\geq1}|\bra{\Psi_0}|M_L(q)|\ket{\Psi_0}|^2\,,
\end{align}
where the notation $\langle \psi_0|\left|O\right||\psi_0\rangle$  indicates the reduced matrix element of a generic operator $O$, which by construction is independent from  $M_A$, $M_A'$, and the direction of $\bmq$. 
The operators $C_L$ and $M_L$ are the charge and the magnetic multipole operators, respectively, with defined angular momentum $L$. Specifically, the longitudinal form factor receives contributions only from the charge multipoles associated with the charge operator, and therefore encodes information about how the electric charge is distributed within the nucleus--hence it is referred to as the charge form factor. 
The transverse form factor, instead, receives contributions from the magnetic multipoles, 
and provides information about the magnetic charge distribution inside the nucleus, and is for this reason referred to as the magnetic form factor.
The charge form factor probes global properties of the nuclear charge distribution, whereas the transverse form factor, built from magnetic multipoles with a non-trivial spin and angular momentum dependence, is sensitive to single-particle and cluster structures of the wave function. Another set of multipoles,  known as electric multipoles $E_L$, can also be defined and, in general, contribute to electromagnetic processes. However, since we are considering only elastic scattering,  these multipoles are forbidden because they violate time-reversal~\cite{Forest1966} (\textit{i.e.}, the invariance of the laws of physics when going forward or backward in time). 

These expressions are very elegant, but how do they help us to understand the structure of the nucleus? 
Multipole operators have a definite angular momentum; therefore, they must satisfy the angular momentum selection rule implied by $0\leq L \leq 2J_A$. Moreover, multipoles have a well-defined parity, namely ${\cal P}(C_L)=(-)^L$ and ${\cal P}(M_L)=(-)^{L+1}$. Since, for elastic scattering, the parity of the initial and final states is the same, and the total parity must be conserved, only even $L$ values contribute to the charge form factor and only odd $L$ values contribute to the magnetic form factor. Let us consider a concrete example. The ground state of $^7$Li has total angular momentum and parity $J^\pi=3/2^+$. The allowed angular momentum for the multipoles are therefore $L=0,1,2,3$. Due to parity conservation, only multipoles $C_0$ and $C_2$ contribute to the longitudinal form factor, while only multipoles $M_1$ and $M_3$ contribute to the transverse form factor. As shown in Figure~\ref{fig:ffactor7li}, these are the only multipoles allowed by the theory and indeed explain the elastic-scattering data.

In the \textit{ab initio} approach of Ref.~\cite{Chambers-Wall:2024fha,Chambers-Wall:2024uhq,King:2024jiq}, matrix elements of the charge and current operators are computed. These are related to reduced matrix elements of the multipole operators via~\cite{Carlson:1997qn}:
\begin{align}
    \langle \psi_0(M_A=J_A)| \rho(\bmq)|\psi_0(M_A=J_A)\rangle&=\sum_{L=0}^{\infty} \sqrt{4 \pi} i^L c_{L J_A} P_L(\cos \theta_q)\langle \psi_0\left\|C_L(q)\right\|\psi_0\rangle,\label{eq:rho_mex}\\
\langle \psi_0(M_A=J_A)| j^y(\bmq)|\psi_0(M_A=J_A)\rangle&=\sum_{L=0}^{\infty} \sqrt{4 \pi} i^{L+1}\frac{c_{L J_A}}{\sqrt{L(L+1)}} P_L^1(\cos \theta_q)\left\langle \psi_0\left\|M_L(q)\right\| \psi_0\right\rangle,\label{eq:j_mex}
\end{align}
where $c_{LJ_A}=\bra{J_A J_A J_A -J_A} L 0\rangle$ is a standard Clebsch-Gordan coefficient, $P_L$ are  Legendre polynomials, $P_L^1$ are associated Legendre functions, and $\theta_q$ is the angle between the quantization axis of the nucleus (\textit{i.e.}, where the nucleus total angular momentum is aligned) and the direction of the exchanged photon, $\hat \bmq$.
In the left hand side of these equations we have the matrix element of the charge and the current operators, sandwiched between the nuclear wave functions with their total angular momentum along the $z$-axis equal to $J_A$. In  {\it ab initio} calculations, these matrix elements are computed directly as a multidimensional integral over the variables describing the nucleons. The reduced matrix elements are then extracted using Eqs.~\eqref{eq:rho_mex} and~\eqref{eq:j_mex}. Let us again look at a concrete example using the $^7$Li transition form factor. As we have seen, in this case we have two independent magnetic multipoles $M_1$ and $M_3$.  Selecting the $z$-axis as the orientation of the  total angular momentum of the nucleus, we can select two independent directions for the exchanged photon, for example  $\theta_q=90^\circ$ (photon along the $x$-axis) and $\theta_q=45^\circ$ (photon $45^\circ$ above the $x$-axis). In this case the linear equations we obtain from Eq.~\eqref{eq:j_mex} are
\begin{align}
 \left\langle \psi_{^7Li}(3/2) \left|{j}^{y}(q\hat{x})\right| \psi_{^7Li}(3/2) \right\rangle
  &=\sqrt{\frac{3\pi}{10}}\left[\sqrt{3} M_1 +\frac{1}{2\sqrt{2}}M_3\right]\,,\\
  \left\langle \psi_{^7Li}(3/2) \left|{j}^{y}\left(q\frac{\hat{x}+\hat{z}}{\sqrt{2}}\right)\right| \psi_{^7Li}(3/2)\right\rangle
  &=\sqrt{\frac{3\pi}{20}}\left[\sqrt{3} M_1 -\frac{3}{4\sqrt{2}}M_3\right]\,.
\end{align}
from which $M_1$ and $M_3$ can be easily extracted. Similar equations can be obtained for the reduced matrix elements $C_0$ and $C_2$ (see~\cite{King:2024jiq} for the explicit derivation). 

The multipole expansion is also computationally more  advantageous. In the $^7$Li example, if we were to compute the matrix elements needed for the transverse form factor in Eq.~\eqref{eq:ff_t} directly, we would need to calculate $M_A\times M_A =4\times 4=16$ matrix elements while using Eq.~\eqref{eq:j_mex} we only need to compute two.
More generally, if $N$ charge (magnetic) multipoles are required, then $N$ independent linear equations must be constructed. Since the direction of $\bmq$ is an independent variable, it is sufficient to evaluate matrix elements for $N$ different directions of $\bmq$ to obtain $N$ linear independent equations. Since $N<M_A^2$, the computational advantage of the multipole expansion is evident.

\begin{figure}
    \centering
    \includegraphics[scale=0.7]{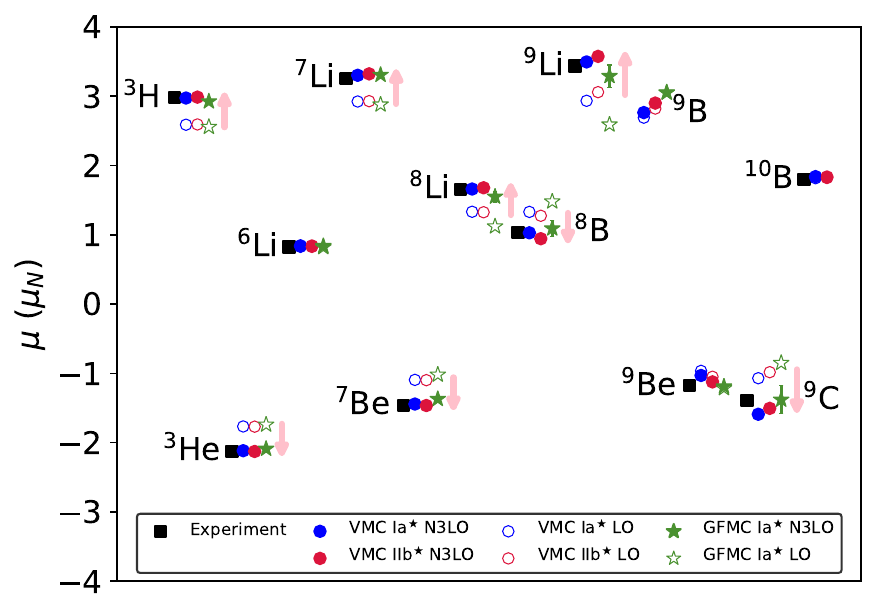}
    \caption{Magnetic moments of selected light nuclei computed using various chiral interactions without (empty symbols) and with (full symbols) two-body currents. The black square represent the experimental value. The circles indicate the calculations performed using variational Monte Carlo (VMC) while the stars the one performed using  Green's function Monte Carlo (GFMC). Figure taken from Ref.~\cite{Chambers-Wall:2024fha}.}
    \label{fig:mag_mom}
\end{figure}

\subsection{Electromagnetic Form Factors and the Shape of Nuclei}

To illustrate the connection between nuclear structure and electron scattering observables, we present results for the charge and magnetic form factors of light nuclei. 
We start with the $^7$Li case, for which we show in Figure~\ref{fig:ffactor7li} the computed transverse (left panel) and longitudinal (right panel) form factors and compare them  with experimental data~\cite{Chambers-Wall:2024fha,Chambers-Wall:2024uhq,King:2024jiq}.
The first thing to notice is the different size of the two contributions: the transverse form factor is $\sim 2$ orders of magnitude smaller than the longitudinal one for $q\lesssim500$ MeV/c, beyond which they become comparable. This highlights how difficult the experimental extraction of the transverse component from the elastic cross section data can be (see Eq.~\eqref{eq:xsec_elastic}).    
The full calculations that contain the current contributions up to N3LO (red lines) reproduce the experimental data well, demonstrating the accuracy with which the chiral interactions and currents describe the electric and magnetic structure of the nucleus. The role of two-nucleon currents is also highlighted: calculations that retain only one-body terms (blue lines) fail to reproduce the data, especially for the transverse form factor, where two-body contributions reach up to $\sim 30\%$ across the momentum-transfer range. For the longitudinal form factors the two-nucleon corrections are very small, except in the very high momentum region.

We now examine the multipole structure of $^7$Li to have a better understanding of its shape. The electromagnetic form factors are given by the sum of two multipoles: $M_1$ (dashed line) and $M_3$ (dotted line) for the transverse form factor and $C_0$ (dashed line) and $C_2$ (dotted line) for the longitudinal one. The $M_1$ multipole exhibits a deep  (or diffraction point) around $q\sim300$ MeV/c.  This feature is indeed characteristic of several p-shell nuclei. In nuclei where higher order multipoles are allowed, this dip is filled by these higher multipoles--just as the $M_3$ does in $^7$Li---generating a smoothly decreasing magnetic form factor. This indicates that the magnetic charge that is closer to the core of the nucleus (recall that large $q$ corresponds to small distance) is mostly generated by octuple magnetic configurations, while the long range part is generated by dipoles.
In a similar fashion, electric charges prefer to be organized as monopoles, far away from the core, where the $C_0$ term dominates. Closer to the core, the term $C_2$ becomes dominant, indicating a quadrupole structure of static electric charges.

Another instructive example is $^{10}$B (see Figure~\ref{fig:ffactor10b}), a $J^\pi=3^+$ nucleus. The allowed multipoles are $M_1$, $M_3$ and $M_5$ for the transverse form factor, and $C_0$ and $C_2$ for the charge form factor. In this case,  $M_3$ is suppressed, while  $M_5$ becomes dominant at short range (large momentum), indicating that the magnetic charge distribution has an hexadecapole component corresponding to a spherical harmonic of angular momentum five. While the presence of the $M_5$ component in the magnetic form factor was well known~\cite{Donnelly:1984rg}, its shape and relative contribution were not determined until the recent \textit{ab initio} calculations of Ref.~\cite{Chambers-Wall:2024uhq,Chambers-Wall:2024fha}--accessing higher magnetic multipoles such as $M_5$ requires that the wave function includes sufficient correlations to support the corresponding multipolarity.
 
In this case, two-body currents (difference between the blue and red lines) are negligible for both charge and magnetic form factors, in contrast to $^7$Li. The reason is that in $^{10}$B only the isoscalar electromagnetic current contributes, \textit{i.e.}, the part of the current that does not distinguish between  protons and neutrons. Comparing the magnetic form factors of $^7$Li and $^{10}$B shows that two-body correlations are much larger in processes that involve the exchange of charges between neutron and protons, or in other words, that neutron-proton correlations are stronger than neutron-neutron and proton-proton ones.

\subsection{Electromagnetic Charge and Moments}

The charge and magnetic form factors have very different behaviors in the low-$q$ limit. The charge form factor reaches its maximum at $q=0$, where its value reduces to the total nuclear charge $Z$, $$F_L(q=0)=Z\,.$$ This follows directly from the leading-order one-body charge operator, which at $q=0$ simply counts the number of protons. Higher-order operators do not contribute at $q=0$. The total charge $Z$  can then be regarded as a global property of the nucleus and serves as a natural normalization of the charge form factor.

The transverse form factor vanishes as $q \rightarrow 0$, as seen in Figures~\ref{fig:ffactor7li} and~\ref{fig:ffactor10b},  which follows
from the $M1$ multipole being $\propto q$ in this limit. 
The magnetic moment $\mu$ is defined as the expectation value of the static $M1$ operator in the nuclear ground state~\cite{Carlson:1997qn} and  it can be extracted from the transverse form factor by taking the limit
\begin{equation}
     \mu=2m_N\lim_{q\rightarrow 0}\frac{F_T(q)}{q}\,,
\end{equation}
where the  $2\,m_N$, with $m_N$ the nucleon mass, is a conventional factor required to express the magnetic moment in units of nuclear magneton, $\mu_N=e/2m_N$~\cite{Carlson:1997qn}. 

As an example, in Fig.~\ref{fig:mag_mom}, we present the magnetic moments of selected light nuclei~\cite{Chambers-Wall:2024fha,Chambers-Wall:2024uhq} obtained using both VMC and GFMC methods. The magnetic moments of $^2$H, $^3$He and $^3$H coincide with the experimental values (filled black squares) by construction, having been used to determine the LECs appearing in the nuclear two-body currents~\cite{Schiavilla:2018udt,Gnech:2022vwr}. All remaining magnetic moments are  predictions. 
Nuclear magnetic moments cannot be explained by simply summing the individual magnetic moments of the constituent neutrons and protons. This single-particle picture (blue and red circles and green stars) fails to reproduce the experimental data, and two-body currents (filled red and blue circles and filled green stars) are essential to agree with the experiment. This has been confirmed by recent {\it ab initio} studies extending into the medium-mass sector~\cite{Miyagi:2023zvv,Martin:2023dhl,Gnech:2023prs,Muller:2024ebw,Zhou:2026hya}.

The magnetic moment can be extracted from electron scattering data, but is often measured with much higher precision in atomic physics experiments~\cite{stone_2014_6a4dw-fmh87}. In either case, the experimental magnetic moments provide a stringent test of {\it ab initio} nuclear theory, as they probe the quality of the nuclear wave functions and the role of two-body current operators.

\subsection{Electromagnetic Radii}

Having discussed the overall shape of the charge and magnetic distributions and their low-$q$ properties, we now turn to their spatial extent, described by the electric $r_E$  and magnetic $r_M$, radii. 
These quantities can be obtained from the low-$q$ expansion of the form factors as:
\begin{align}
\frac{1}{Z}F_L(q) &\approx 1 - \frac{1}{6}r_E^2q^2 + \mathcal{O}(q^4)\, ,\label{eq:re}\\
    \frac{2m_N}{q\,\mu}F_T(q) &\approx 1 - \frac{1}{6}r_M^2q^2 + \mathcal{O}(q^4)\label{eq:rm}\, .
\end{align}
As suggested by the equations above, the radii can be extracted directly from electron scattering data. Complementary determinations come from high-precision atomic spectroscopy experiments, whose accuracy is increasingly challenging the nuclear theory community to compute these observables at comparable precision~\cite{Filin:2020tcs,Yang:2025rcx,Sun:2026eep}. Electromagnetic radii thus provide important benchmarks for {\it ab initio} nuclear theory.

\begin{figure}
    \centering
\includegraphics[scale=0.48]{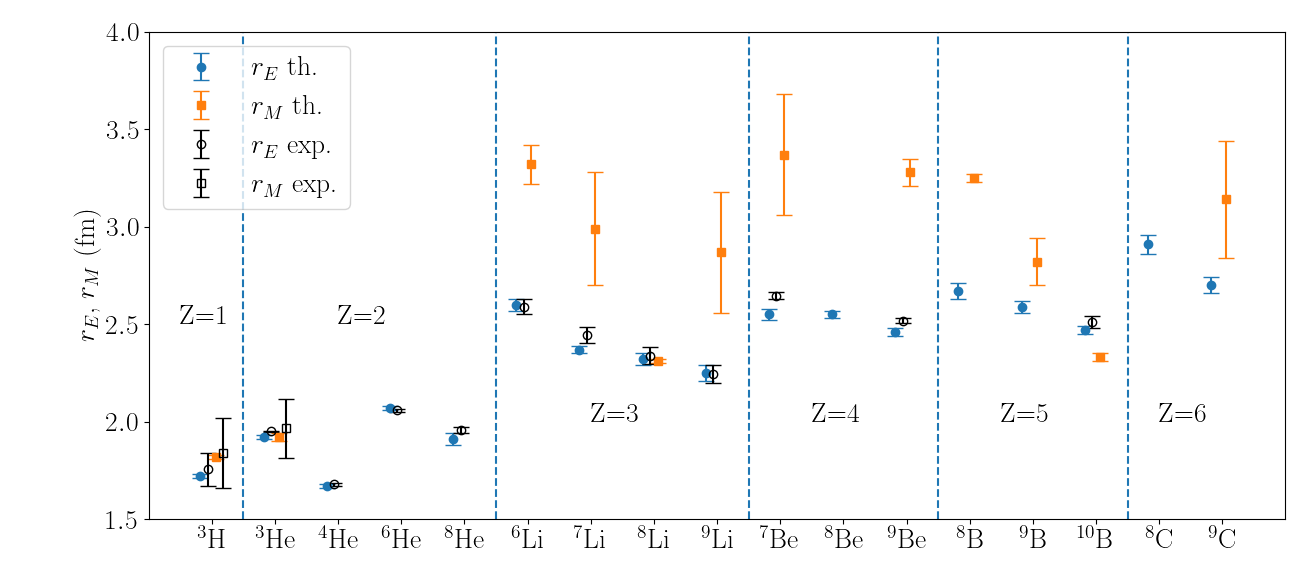}
    \caption{Charge (blue dots) and magnetic (orange squares) radii of selected light nuclei, extracted from the corresponding form factors computed in the $\chi$EFT framework, using the VMC approach~\cite{King:2025akz}. The empty black circle (squares) are the experimental results for the charge (magnetic) radii. The experimental magnetic radii are available only for $^3$H and $^3$He.}
    \label{fig:radii}
\end{figure}

As an example, Fig.~\ref{fig:radii} presents the charge (filled blue dots) and magnetic (filled orange squares) radii of Ref.~\cite{King:2025akz}, extracted from the charge and magnetic form factors, compared with the available experimental data (empty symbols). 
The figure is organized by isotopic chains.
While charge radii are experimentally well determined across the light nuclei shown, magnetic radii suffer from a much sparser experimental coverage. New measurements of electromagnetic radii--including from high-precision atomic spectroscopy--would be extremely valuable in constraining nuclear models and cross-validating our understanding of nuclear structure.

\section{Conclusions}\label{sec:conclusions}

Elastic electron scattering provides a direct window into how electric and magnetic charges are distributed inside nuclei. The cross section can be written in terms of the longitudinal and transverse form factors that encode this information. In particular, the longitudinal form factor depends explicitly on the charge operator and therefore gives an indication on how the charge is distributed inside the nucleus. The transverse form factor, instead,  probes the nuclear magnetization distribution, arising from the currents and spins of the constituent nucleons.
To study these distributions, we perform a multipole decomposition of the form factors. This reveals a rich structure that evolves with the momentum transfer or, equivalently, with the spatial resolution of the probe. At low momentum, probing large distances, the nucleus appears as nearly spherical distribution of charge, with a dipole magnetization distribution. As the momentum increases, higher multipoles emerge according to the nuclear spin and parity selection rules, revealing quadrupole, octupole, and higher-order structures in both the charge and magnetization distributions.
Studying the low-momentum behavior of the form factors allows us to define fundamental properties of the nucleus, including electromagnetic moments and radii. We presented examples of calculated observables in light nuclei using {\textit{ab initio}} methods and highlighted the role of many-nucleon correlations and currents in reproducing the experimental data.

While electromagnetic form factors provide a detailed picture of the nuclear charge and magnetization distributions, a complete understanding of matter distributions within the nucleus requires complementary electroweak probes, including, neutrino and parity-violating electron scattering experiments. Future progress in this field will rely on the interplay between new experimental measurements--spanning electron scattering and high-precision atomic spectroscopy--and continued advances in {\it ab initio} theory across a wider range of nuclei and momentum scales. These developments are essential not only for nuclear structure physics, but also for our understanding of astrophysical processes, neutrino--nucleus interactions, and tests of fundamental symmetries.

\section*{Acknowledgments}
This work is supported by the US Department of Energy under Contracts No. DE-SC0021027 (G.~C.-W., S.~P.), DE-AC05-06OR23177 (L.~A., A.~G). G.~C.-W. acknowledges support from the NSF Graduate Research Fellowship Program under Grant No. DGE-213989. G.~B.~K. acknowledge support  from the Laboratory Directed Research and Development program of Los Alamos National Laboratory under project number 20240742PRD1. We thank the Nuclear Theory for New Physics Topical Collaboration, supported by the U.S.~Department of Energy under contract DE-SC0023663, for fostering dynamic collaborations. A.~G. acknowledges the direct support of the Nuclear Theory for New Physics Topical collaboration. 

\bibliographystyle{unsrt}
\bibliography{bibs}

\end{document}